# Parametric Resonance Enhancement in Neutron Interferometry and Search for Non-Newtonian Gravity

Vladimir Gudkov

Department of Physics and Astronomy, University of South Carolina,

Columbia, SC 29208

H. M. Shimizu

High Energy Accelerator Research Organization (KEK),

1-1 Oho, Tsukuba, Ibaraki 305-0801, Japan

Geoffrey L. Greene

Department of Physics, University of Tennessee, Knoxville, TN, 37996, and

Physics Division, Oak Ridge National Laboratory, Oak Ridge, TN, 37831

## Abstract

The parametric resonance enhancement of the phase of neutrons due to non-Newtonian anomalous gravitational is considered. The existence of such resonances are confirmed by numerical calculations. A possible experimental scheme for the observation of this effect is discussed based on an existing neutron interferometer design.

## Introduction

Experimental searches for anomalous gravity have been carried out for many years (see Adleberger, *et al.*[1] for a comprehensive review). More recently, this effort has been stimulated by theoretical models with a mechanism of gravitational unification based on a compactification of extra dimensions which leads to existence of massive gravitons (see, for example [2,3,4,5] and references therein). Neutrons, as a neutral, heavy, and (almost) stable particle, are an attractive probe for anomalous gravitational forces with ranges from nuclear sizes up to a macroscopic scale. Indeed, low energy neutron scattering and ultra cold neutron experiments have been used to set some of the best constraints on the possible magnitudes of anomalous gravity at short range.[6,7,8,9,10,11].

In a recent paper [12] a method to search for exotic gravitational interactions using neutron interferometry was proposed. In the proposed method, two thin plates of a dense material are placed in one path of the interferometer. The plates, while thin, are assumed to be thicker than the supposed range of the anomalous interaction. One of the plates is fixed. The other is attached to a translation slide which allows a motion along the direction of the neutron path. Due to nuclear effects, the insertion of the plates introduces a phase change between the paths. The presence of an anomalous short range interaction will introduce another which is, in principle, experimentally distinguishable due to its dependence on the plate separation. To analyze this system, we assume a hypothetical, short range non-Newtonian interaction whose form is given by[13]

$$V(r) = -\frac{GMm}{r}(1+\alpha e^{-r/\lambda}), \qquad (1)$$

where G is the gravitational constant, $m$ and $M$ are interacting masses, $\alpha_G$ is a dimensionless parameter and $\lambda$ is effective range of gravitational interactions. The potential for a neutron propagated over a plate of matter of uniform density $\rho$, as a function of $x$, the distance from the surface is given by[14],

$$V_{eff} = 2\pi G \alpha_G m_n \rho \lambda^2 e^{-x/\lambda} \qquad \text{outside the material;}$$

$$V_{eff} = 2\pi G \alpha_G m_n \rho \lambda^2 (2 - e^{-x/\lambda}) \qquad \text{inside the material.}$$

If the separation between the plates varies from $L=0$ to $L \geq \lambda$, where $\lambda$ is the range of the anomalous gravity, the phase of the neutron will change. This phase shift has been calculated for thermal neutrons [12]. It has been noted [12] that the form of this phase shift suggests the possible parametric resonance enhancement of the gravitation interactions during neutron propagation through the slabs. While very small, this enhancement may be observable. In this paper we examine the conditions under which the parametric phase resonance may enhance the manifestations of anomalous gravity and discuss a potentially realizable experimental arrangement,

## Discussion

We consider the general solution of Schrödinger's equation for neutron propagation through two parallel plates of the same material separated by distance $L$, assuming only that the size of the plates are much larger than both the distance $L$ and the range of the short gravitational interaction $\lambda$. In the absence of the gravitational interactions, neutron is exposed only to Fermi pseudo-potential due to coherent nuclear interactions inside the plates. This process corresponds to a solution of one-dimensional Schrödinger equation for a square potential well/barrier of range $L$ with a potential equal to the Fermi potential

$V_F = \dfrac{2\pi\hbar^2}{m_n} Nb$, where $N$ is a number of nuclei in a unit volume, and $b$ is a coherent neutron scattering length. If one defines the neutron wave number in the material as

$k = \sqrt{\dfrac{2m_n}{\hbar^2}(E_n - V_F)}$ and in vacuum as $k_0$, one can write the transmission coefficient for neutron propagation through the vacuum space between two plates as:

$$T_0 = \dfrac{2k_0 k e^{-ikL}}{2k_0 k \cos(k_0 L) - i(k^2 + k_0^2)\sin(k_0 L)} . \qquad (2)$$

The argument of this expression gives the phase shift of the neutron wave.

If a short range gravitational interaction exists, the square potential well/barrier will be modified. Choosing the origin of the reference frame for neutron propagation along axis $x$ at the surface of the first plate, the modification will lead to the changes of neutron wave numbers in the first plate of:

$$k^2 \rightarrow k^2 + 2a^2 - a^2 e^{(x-d)/\lambda}\left[1 - e^{-L/\lambda}\right]; \qquad (3)$$

between the plates as:

$$k_0^2 \rightarrow k_0^2 + a^2(e^{-x/\lambda} + e^{(x-L)/\lambda}); \qquad (4)$$

and, in the second plate as:

$$k^2 \rightarrow k^2 + 2a^2 + a^2 e^{(d-x)/\lambda}\left[e^{L/\lambda} - 1\right]; \qquad (5)$$

where $a^2 = 2\pi G \alpha_G m_n \rho \lambda^2$. This leads to the superposition of exponential and constant potentials in corresponding Schrödinger equations. To obtain a transmission coefficient in the presence of anomalous gravitation, it is necessary to solve these equations for each region and to match the solutions. The solutions of Schrödinger equations inside the slabs can be represented in terms of Bessel functions. Between the slabs, the potential is proportional to the hyperbolic cosine. This leads to Mathieu's modified differential equations which can be solved analytically in terms of Mathieu functions (rather than as exponential functions for the case of the square-well potential). The complete analytic expression for the transmission coefficient is quite long and is not provided here.

The fact that neutron transmission coefficient results from a solution of a Mathieu - type equation (which is often used to describe parametric resonances) suggests the possibility of observing an enhancement in neutron propagation provided the proper resonance conditions are satisfied.

Before applying of our results to the search for anomalous gravity, let us consider some general issues related to the possibility of existence of parametric resonances (PR) in one-dimensional transmission process. We first note that a similar phenomenon has been discussed recently for the coherent propagation of neutrinos through materials with a variation of density profile (see, for example papers [15] and references therein). The PR enhancement of neutrino oscillations has been calculated for a variation of density of material over only "one and a half" period of density modulation[16]. These calculations demonstrate that PR can exist in transmission processes without a requirement of "pure" periodic structure of density profile variations.

It should also be noted that the possibility of existence of PR in one dimensional transmission process has been known for a long time. To our knowledge, L. P. Pitaevsky (see [17] and references therein) first pointed out by that the solution of the one-dimensional Schrödinger equation for harmonic oscillator exactly coincides the solution the one-dimensional potential barrier propagation problem (provided that the transmitted particle momentum $k(x)$, as a function of distance, is replaced by the frequency of harmonic oscillator as a function of time in the Schrödinger equation). The correspondence of the solutions for these two problems (harmonic oscillator and transmission) has been studied [17] for the case of exactly solvable Eckart's potential [18], which analytically exhibits the existence of quantum parametric resonance (quantum parametric amplifier [17]). Following this approach, we analyze conditions for the existence of PR in the case of neutron transmission in the two slabs framework with short range gravity.

The potential in between the slabs, given by Eq.(4), can be written as

$$k_0^2(1+\eta\cosh(x/\lambda)), \qquad (6)$$

where $\eta = 2a^2 \exp(-L/2\lambda)/k_0^2$. Taking into account the identity

$$\cosh(x/\lambda) = \frac{\sinh(L/\lambda)}{(L/\lambda)} + \sum_{n=1}^{\infty}\left(\frac{2(-1)^n(L/\lambda)\sinh(L/\lambda)}{(L/\lambda)^2 + n^2\pi^2}\cos\left(\frac{n\pi x}{L}\right)\right), \qquad (7)$$

one sees that the quantum parametric resonance appears when

$$2k_0 / \left(1+\eta\frac{\sinh(L/\lambda)}{(L/\lambda)}\right) = \frac{n\pi}{L}. \qquad (8)$$

For very small parameter $\eta$ ($\eta\sinh(L/\lambda)/(L/\lambda) \ll 1$), which is almost the case for gravitational interactions, this can be simplified as

$$\lambda_n \simeq 4L/n. \qquad (9)$$

(Here $\lambda_n = 2\pi/k_0$ is neutron wavelength, which is not to be confused with the range of gravitational interaction $\lambda$. ) This leads to an estimate for the resonance width as

$$\gamma \simeq \frac{a^2 \lambda_n^2}{\pi^2} \frac{(L/\lambda)\sinh(L/\lambda)}{(L/\lambda)^2 + 16\pi^2(L/\lambda_n)^2} e^{-L/(2\lambda)}. \tag{10}$$

(Recall that $a^2$ is proportional to $\lambda^2$, as can be seen from the definition of $a^2 = 2\pi G \alpha_G m_n \rho \lambda^2$.) From this analysis, one concludes that, to be able to observe this resonance in practice, one must work with rather long neutron waves. This condition provides the challenge to the observation of PR using thermal neutrons [16].

We emphasize that we are dealing with a resonance enhancement of the *phase* of transmission coefficient for neutron propagation. This particular subject has not been explicitly considered in the literature, which has addressed problems related to particle *transmission* in one dimensional barrier problems. However, the revision of a number of well-known solutions of Schrödinger equation for the transmission problem indicates the possibility of resonance phase enhancement in one-dimensional scattering/transmission process. For example, it has been shown [19] that the transmission amplitude for exponentially decreasing potentials have an infinite number of singularities in the complex momentum plane. The position of the singularity nearest to the real axis is defined only by the slope of the potential and not by it's value. For a pure exponential potential, the nearest singularity lies on the imaginary axis. It was pointed out [19] that for the case of two overlapping potentials, corresponding to the situation considered in this paper, the transmission amplitude has a second order pole at the same position defined only by the slope, $k = -2i/\lambda$. This implies that when neutron wavelength $\lambda_n$ is comparable with the scale of (anomalous gravitational) interaction $\lambda$, the phase of the transmission amplitude is most sensitive to this interaction. It is important that the minimal distance of the singularities from the real axis is not equal to zero and that the poles are allowed to move off the imaginary axis for non-pure exponential potential. To illustrate the sensitivity of the phase of the transmission amplitude to the slope of the potential, one may consider a "toy" model for neutron transmission above the potential barrier which consists of two potentials: the first is a localized strong potential and the second is a weak exponentially decreasing potential. The amplitude $T$ for neutron transmission through the sum of these potentials is proportional to the product of two transmission amplitudes for each (it can be approximated as a product of these two transmission amplitudes for each potential multiplied by a function which depends on transmission/reflection coefficients for each potential and on geometry of the problem, see for example [20])

$$T \sim t_s t_w. \tag{11}$$

If one potential is very weak compare to other one, its transmission coefficient is very close to one ($|t_w|^2 \simeq 1$), and total transmission is, for all practical purposes, governed by the properties of the strong potential. The phase of $t_w$, however, has an additive contribution to the total phase of the $T$-amplitude

$$Arg(T) = Arg(t_s) + Arg(t_w) + .... \tag{12}$$

and could be as much as

$$Arg(t_w) \sim \tan^{-1}(\pi \lambda_n / \lambda) \tag{13}$$

or

$$Arg(t_w) \sim \tan^{-1}\left(\frac{2(\pi \lambda_n / \lambda)}{(\pi \lambda_n / \lambda)^2 - 1}\right) \tag{14}$$

for the first and second order poles, correspondingly. This implies that by going to a very low neutron energy region, in order to be close to the nearest pole, the phase of neutron transmission coefficient will be sensitive to small exponentially decreasing potentials. This rather surprising observation is confirmed by cases with exact analytical solutions for transmission through potential barriers with exponentially dependent potentials (see, for example, Eckart's potential [18], and number of other solved problems [21] [22]). For all these cases, the transmission amplitudes are meromorphic functions with the positions of the poles depended on the value of exponential slopes but not on the value of the potentials themselves.

Consideration of such toy-models suggests that an enhanced sensitivity to gravitational potential for the phase of the transmission coefficient will occur when neutron wavelengths are comparable to the scale of anomalous gravitational interactions and to the characteristic distance between two slabs in the interferometric experiment. This condition is consistent with the requirements for existence of parametric resonance. It is important to note that the formulae discussed above were obtained using a number of simplifications to allow compact analytical expressions. They can be used only to provide a qualitative understanding of the process. Realistic estimates of the magnitude of possible manifestations of the PR must follow from exact numerical calculations [16], in the low energy range, where observable effects are anticipated. For example, in calculating a phase shift for two slabs experimental setup [16] with $\lambda = 5nm$, and the distance $L = 10nm$ as a function of neutron wavelength $\lambda_n$ (see Fig.1), one observes a very narrow resonance for $\lambda_n \simeq 375.5 \overset{\circ}{A}$. Numerical calculations confirm that the position of the resonance practically does not depend on intensity of gravitational interaction.

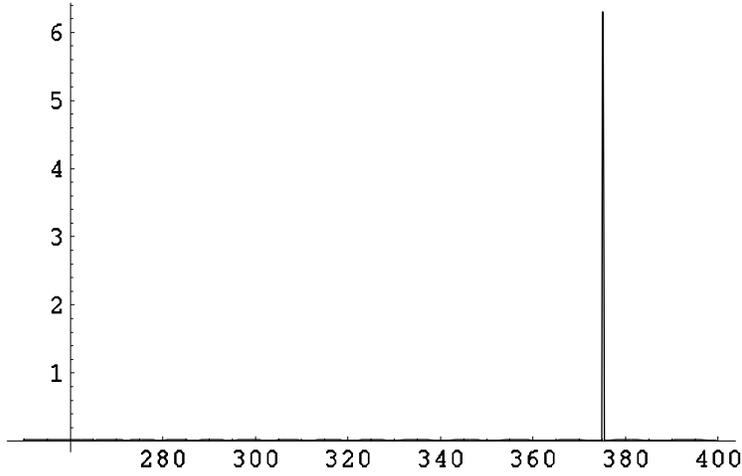

Fig. 1: Gravitational phase as a function of neutron wavelength.

As a demonstration of the sensitivity of the method, we calculated excluded regions (white and black) for the discussed setup for $\lambda_n = 300 \, \overset{\circ}{A}$ assuming the experimental sensitivity to the phase at the level $10^{-3}$ *rad* (see Fig. 2).

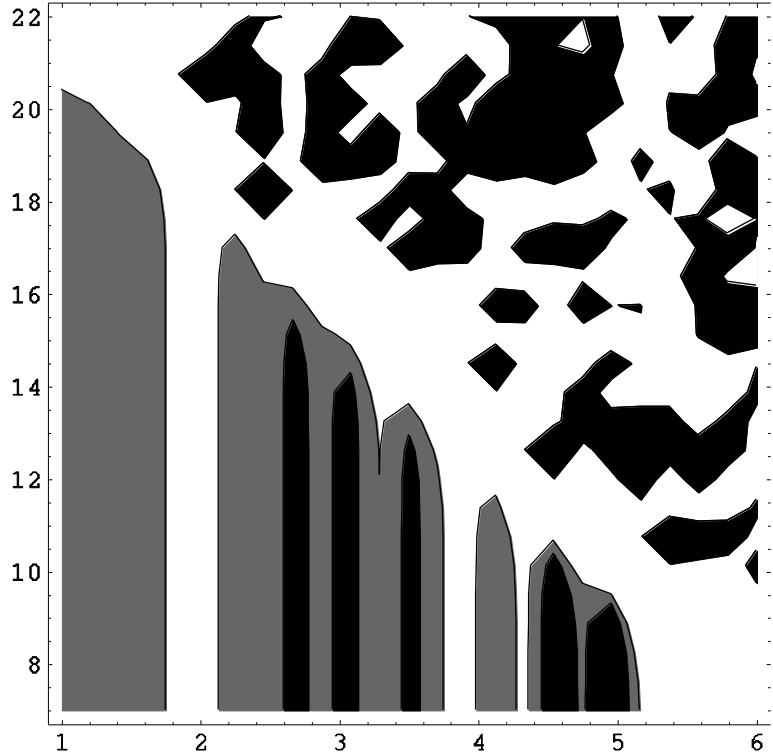

Fig. 2: Excluded (white and black) region for anomaly gravitational interaction for 300 angstrems neutron wavelength and experimental sensitivity of 0.001 *rad* as logarithm of intensity of interaction versus of the logarithm of the scale of the interaction in *nm*.

These numerical results confirm the simplified analysis above and provide the first detailed exploration of phase parametric resonance in neutron interferometry. We note that, although the effect suggested by the above analysis are very small, they may be accessible by using a new type of neutron interferometer. In the next section we discuss a possible experimental realization.

## Experimental Realization

To date, most neutron interferometry experiments have been performed using perfect silicon interferometers [23]. Such devices are dependent on Bragg diffraction are thus limited to neutrons having wavelength on the order of *0.2nm* or less. For reasons of experimental practicality we assume that we will be limited to slab spacings (and thus neutron wavelengths) of $L \geq 1 nm$. Kitaguchi et al. successfully demonstrated neutron spin interferometry at $\lambda = 0.88\ nm$ [24]. They used artificial multilayers to configure a Jamin-type interferometer. This technique offers the prospect of extension to the very-cold-neutron (VCN) regime with neutron wavelengths of several nanometers since, unlike perfect crystals, the lattice constant of the multilayers can be varied. In the following, we estimate the feasibility of using a Kitaguchi interferometer at $\lambda = 10\ nm$ to search for an anomalous gravitational phase change as described above.

For simplicity, we consider the case shown in Fig. 3. The etalons are separated by distance *D* and the neutron beam is incident to the etalon pair at angle θ. Let *z* be the beam axis as *z*-axis, *x* be the horizontal, and *y* the vertical axis. For $\lambda = 10\ nm$, the neutron velocity along *z*-axis is $v_z = 40\ m\ s^{-1}$. Each etalon is assumed to me constructed precision glass flats with dimension $w_E \times w_E \times t_E$ and a gap of *G*. Multilayer mirrors with the effective lattice constant of *d* are deposited on the central square of $w_E \times w_E$. θ is chosen to satisfy $\sin\theta = h/(2mv_z d) = 0.21$, where *h* is Planck's constant and *m* the neutron mass.

Using the parameters of existing Jamin-type neutron interferometers, we assume that

$w_E = 4.2 \times \sqrt{\pi}/2 \ cm = 3.7 \ cm$, $w_M = 2.0 \times \sqrt{\pi}/2 \ cm = 1.8 \ cm$, $d = 24 \ nm$ \$d=24, $G = 9.75 \times 10^{-4} \ cm$ and $D = 34 \ cm$.

The spatial *x*-width of a mirror seen from the beam is given by $w_x = w_M \sin\theta - G\cos\theta = 0.37 \ cm$. The beam path separation in the interferometer is given by $W_x = 2G\cos\theta = 1.9 \times 10^{-3} \ cm$. We use $W_x$ for the effective spatial *x*-width so that we can measure the relative phase change between two separated beam paths transmitting material plates with a small gap and without a gap.

Figure 4 shows the phase space distribution of the acceptance along *x* and *y*. The acceptance is given as

$$V_{XY} = \left(\frac{v_z W_x w_M}{D}\right)^2 = 1.6 \times 10^{-5} \ cm^2 \ (m/s)^2. \tag{15}$$

To estimate the neutron intensity, we assume the use of the VCN beam port at PF-2 of Institute Laue Langevin Research Reactor $\Phi = 10^5 \ cm^{-2} \ s^{-1} \ (m/s)^{-1}$ [25][26]. We assume the beam divergence is limited by acceptance of a nickel-coated neutron guide, corresponding to $\Delta v_x = \Delta v_y \leq 7 \ m/s$, and obtain $\Phi/(\Delta v_x \Delta v_y) = 2 \times 10^3 \ cm^{-2} s^{-1}(m/s)^{-1}$, where $\Delta v_x$ and $\Delta v_y$ are the beam divergence along *x* and *y* direction, respectively.

The counting rate $N/t$ will be given by

$$\frac{N}{t} = \frac{\Phi}{\Delta v_x \Delta v_y} V_{XY} T^2 \frac{1}{2} \Delta v_z . \tag{16}$$

Making reasonable assumptions about neutron capture in the etalon, we assume a transmission of $T = 0.48$. The $\Delta v_z$ is given as

$$\Delta v_z = v_z \times \frac{\Delta \lambda}{\lambda} = 0.96 \ m/s , \tag{17}$$

And we assume $\Delta \lambda / \lambda = 0.024$.

We obtain

$$\frac{N}{t} = 3.5 \times 10^{-3} \ s^{-1} . \tag{18}$$

Assuming $N = 10^6 / K^2$ corresponds to the phase accuracy of $\Delta \phi = 10^{-3} \ rad$, where $K$ is the visibility of the interferometer. Using the parameters of the current Kitaguchi device we determine $K=0.6$. and find $t=25$ years which is clearly impractical. However, if the air gap of the etalon were expanded up to $G=1$ *mm*, the counting rate is remarkably increased as

$$\frac{N}{t} = 37 \ s^{-1} , \tag{19}$$

which corresponds to $t=21$ hours. Jamin-type interferometer with larger gap etalons are currently under development [24]. In addition, further improvements may be expected by reducing the etalon size to match the beam path separation requirement and by optimizing the effective lattice constant of multilayers, etc. More detailed experimental realization and sensitivity will discussed in a future work.

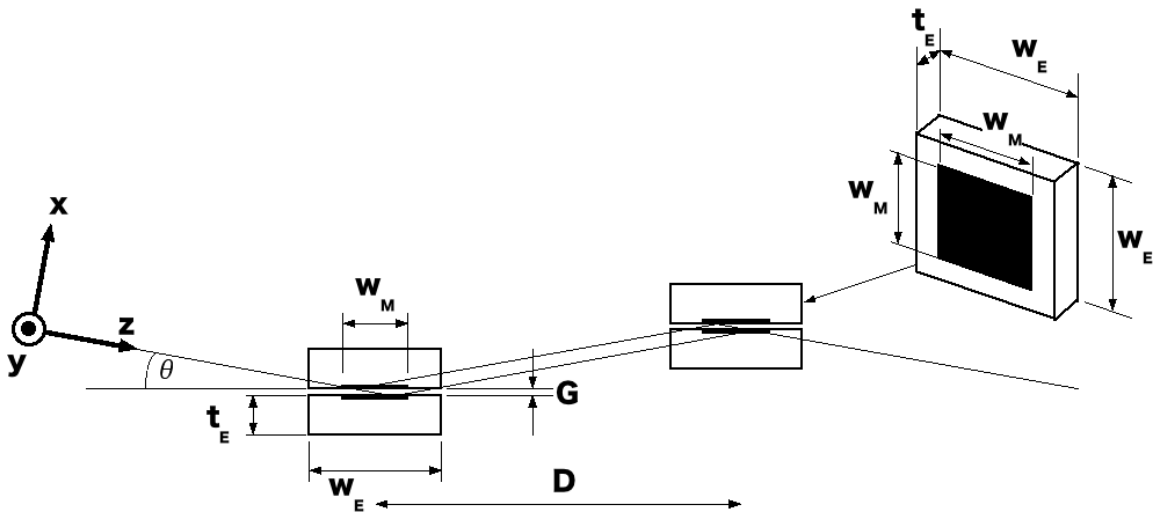

**Figure 3: Schematic view of the Kitaguchi's interferometer.**

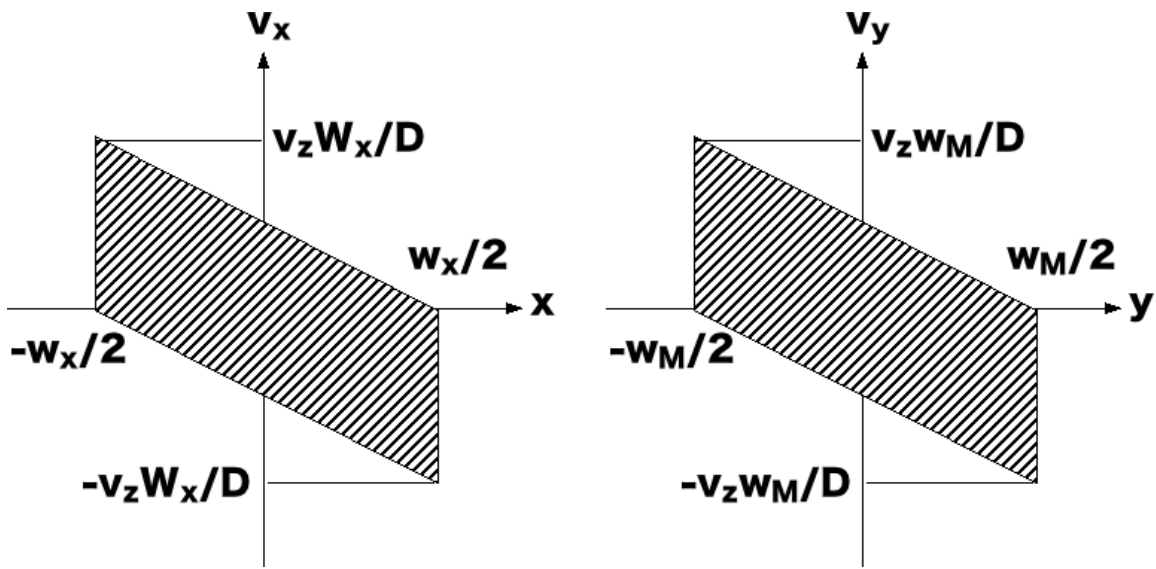

Figure 4: Square-shape etalon used for the estimation.

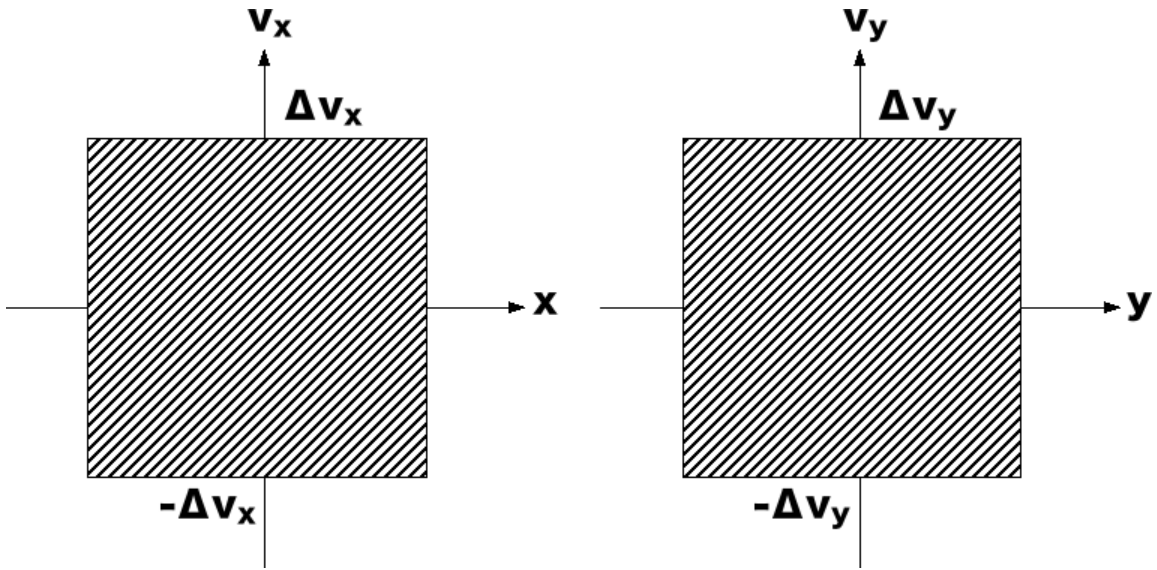

Figure 5: Phase space distribution of incident beam.

**Acknowledgments**

This work was supported by the DOE grants no. DE-FG02-03ER46043 and DE-FG02-03ER41258.

---